\documentstyle[12pt,amsmath,amssymb]{article}

\newcounter{resultcounter} 
\stepcounter{resultcounter}

\newcommand{\D}{{\mathcal D}}
\renewcommand{\Re}{Re}
\renewcommand{\Im}{Im}
\newcommand{\fer}[1]{(\ref{#1})}

\renewcommand{\H}{{\mathcal H}}
\renewcommand{\S}{{\mathcal S}}
\renewcommand{\O}{{\mathcal O}}
\newcommand{\B}{{\mathcal B}}
\renewcommand{\L}{{\mathcal L}}

\renewcommand{\a}{\alpha}
\newcommand{\R}{{\mathcal R}}
\renewcommand{\S}{\Sigma}

\newcommand{\h}{{\mathbf h}}
\newcommand{\F}{{\mathcal F}}

\newcommand{\unit}{{\mathbf 1}}

\newcommand{\M}{{\mathcal M}}
\newcommand{\N}{{\mathcal N}}

\newcommand{\C}{{\mathcal C}}

\newcommand{\ls}{\lambda(s)}
\newcommand{\lsp}{\lambda(s')}

\newcommand{\PV}{{\mathcal PV}}

\renewcommand{\tilde}[1]{\widetilde{#1}}

\begin{document}

\title{On the quasi-static evolution \\ of nonequilibrium steady states}

\author{
Walid K. Abou Salem\footnote{{\it Current address: Department of Mathematics, University of Toronto, M5S 2E4 Toronto, Canada}}\\
Institute for Theoretical Physics, ETH Zurich\\
CH-8093 Zurich, Switzerland\footnote{{\it E-mail: walid@itp.phys.ethz.ch}} }
\date{\;}

\maketitle

\begin{abstract}
The quasi-static evolution of steady states {\it far from equilibrium} is investigated from the point of view of quantum statistical mechanics. As a concrete example of a thermodynamic system, a two-level {\it quantum dot} coupled to several reservoirs of free fermions at different temperatures is considered. A novel adiabatic theorem for unbounded and nonnormal generators of evolution is proven and applied to study the quasi-static evolution of the nonequilibrium steady state (NESS) of the coupled system. 
\end{abstract}


\section{Introduction}

Recently, there has been substantial progress in understanding and rigorously proving the asymptotic convergence (as time $t\rightarrow\infty$) of a state of a thermodynamic system, say one composed of a finitely extended system coupled to one or more thermal reservoir, to a steady state, both in equilibrium ([JP1,2,BFS,FM,M1,2, DJ]) and far from equilibrium ([Ru1,2,JP3,FMUe,MMS1,2]) from the point of view of quantum statistical mechanics. After the state of a certain thermodynamic system reaches a steady state, it is natural to ask how the state will evolve if the system is perturbed slowly over time scales that are large compared to a generic relaxation time of the system, and how much the state of the system will be close to the instantaneous (non)equilibrium steady state. 

This question was first addressed in [A-SF1], where the {\it isothermal theorem}, an adiabatic theorem for states close to thermal equilibrium, has been proven, and applications of this theorem to reversible isothermal processes have been discussed. Here, we pursue this question further by investigating the quasi-static evolution of states {\it far from equilibrium} from the point of view of quantum statistical mechanics. 

According to the spectral approach to nonequilibrium steady states (NESS), the latter corresponds to a zero-energy resonance of the (adjoint of the) C-Liouvillean; (see [JP3,MMS1,2]). Since the C-Liouvillean is generally nonnormal and unbounded, we prove an adiabatic theorem for generators of evolution that are not necessarily bounded or normal. This theorem can be extended to study the adiabatic evolution of quantum resonances. [A-SF2] 

As a concrete example of a thermodynamic system, we consider a system composed of a two-level quantum system coupled to several fermionic reservoirs at different temperatures (for example, a {\it quantum dot} coupled to electrons in several  metals). We apply the general adiabatic theorem to study the adiabatic evolution of the NESS for this system. The main ingredients of our analysis are an adiabatic theorem for nonnormal and unbounded generators of evolution, a concrete representation of the fermionic reservoirs (Araki-Wyss representation [ArWy]), the spectral approach to NESS using C-Liouvilleans, and complex deformation techniques as developed in [HP,JP1,2,3].

The organization of this paper is as follows. In section 2, we state and prove a general adiabatic theorem (Theorem 2.2). This is the key result of this section, which we apply in the subsequent sections to study the quasi-static evolution of nonequilibrium steady states. In section 3, we discuss the concrete physical model we consider: a two level quantum system coupled to several fermionic reservoirs at different temperatures.\footnote{The analysis can be directly generalized to the case when the small system is coupled to several bosonic reservoirs by using methods developed in [MMS1,2].} In section 4, we study the C-Liouvillean corresponding to the coupled system using complex deformation techniques (Theorem 4.3), and recall the relationship between the NESS and a zero-energy resonance of the C-Liouvillean (Corollary 4.4). In section 5, we apply Theorem 2.2 to study the adiabatic evolution of the NESS of the coupled system. The main result of this section is Theorem 5.1. We also remark on the strict positivity of entropy production in the quasi-static evolution of NESS, and on a concrete example of the isothermal theorem [A-SF1]. Some technical details and proofs are collected in an Appendix.

\bigskip

\noindent{\bf Acknowledgements}

\noindent I thank J\"urg Fr\"ohlich, Gian Michele Graf and Marcel Griesemer for useful discussions. I am also grateful to an anonymous referee for a very critical reading of the manuscript and for helpful suggestions. The partial financial support of the Swiss National Foundation during the initial stages of this work  is gratefully acknowledged. 



\section{A general adiabatic theorem}

So far, adiabatic theorems that are considered in the literature deal with generators of evolution which are self-adjoint; (see for example [AE]). This is expected, since the generator of dynamics in quantum mechanics, the Hamiltonian, is self-adjoint. However, for systems out of equilibrium, a generally nonnormal and unbounded operator, the so called C-Liouvillean, can be used to generate an equivalent dynamics on a suitable Banach space. Since we are interested in studying the quasi-static evolution of NESS, it is useful to prove an adiabatic theorem for nonnormal generators of time evolution. This is what is done in this section.

Consider a family of closed operators $\{ A (t) \}, t\in {\mathbf
R}^+,$ acting on a Hilbert space $\H.$ We make the following
assumptions on $A(t)$ in order to prove the existence of a time
evolution and to prove an adiabatic theorem. {\it All} of these 
assumptions will be verified in the applications which are considered in the subsequent sections.

\begin{itemize}

\item[(A1)] $A(t)$ is a generator of a contraction semi-group for all $t\in {\mathbf R}^+.$

\item[(A2)] $A(t)$ have a common dense domain $\D\subset\H$ for all $t\in {\mathbf R}^+.$

\item[(A3)] For $z\in \rho(A(t)),$ the resolvent set of $A(t)$, let $R(z,t):= (z-A(t))^{-1}$. Assume that $R(-1,t)$ is bounded and differentiable as a bounded operator on $\H$, and that $A(t)\dot{R}(-1,t)$ is bounded, where the $(\dot{} )$ stands for differentiation with respect to $t$. Moreover, assume that for every $\epsilon > 0,$ $-\epsilon\in \rho(A(t)).$ 

\end{itemize}

Let $U(t)$ be the propagator that satisfies
\begin{equation}
\partial_t U(t) \psi= -A(t)U(t) \psi\; , \; U(t=0) =1 \; ,
\label{timeevolution}
\end{equation}
for $t\ge 0 ; \; \psi\in\D.$ We have the following result.

\bigskip

\noindent {\bf Lemma 2.1}

{\it Suppose that assumptions (A1)-(A3) hold. Then the
propagator $U(t)$ satisfying (\ref{timeevolution}) exists and is
unique, and $\| U(t)\psi \| \le \|\psi\|,$ for $\psi\in\D.$}
\bigskip

The result of Lemma 2.1 is standard, and it follows from assumptions (A1)-(A3) above and Theorem X.70 in [RS2].\footnote{Choose $\eta > 0$ and let $\tilde{U}(t)$ be the propagator generated by $\tilde{A}(t):=A(t)+\eta.$ It follows from (A1) that $\tilde{A}(t)$ is a generator of a contraction semigroup. Furthermore, for $t,t' \in {\mathbf R}^+$, $\tilde{A}(t')\tilde{A}(t)^{-1}$ is bounded due to the closed graph theorem and (A2) (see [RS1]).
Moreover, for small $|t-t'|$, $|| (t'-t)(\tilde{A}(t')\tilde{A}(t)^{-1}-{\mathbf
1})||=|| \tilde{A}(t)\dot{\tilde{A}}^{-1}(t)||+o(|t-t'|)$, which is bounded due to (A3). By Theorem X.70 in [RS2] (or Theorem 2, Chp XIV in [Yo],
section 4), this implies, together with (A1) and (A2), that $\tilde{U}(t)$ exists and is unique.  In particular, $\|\tilde{U}(t)\psi \|\le 1$ uniformly in $t\ge 0$ (for $\|\psi\|=1$). We also have $\| U(t)\| = e^{\eta t}\|\tilde{U}(t)\|.$ Taking the limit $\eta\rightarrow 0$ gives $\|U(t)\psi\|\le 1.$}

Assume that $A(t)\equiv A(0)$ for $t\le 0$, and that it is
perturbed {\it slowly} over a time $\tau$ such that
$A^{(\tau)}(t)\equiv A(s)$, where $s:=\frac{t}{\tau}\in [0,1] $ is
the rescaled time. The following additional two assumptions are needed to
prove an adiabatic theorem.

\begin{itemize}

\item[(A4)] The eigenvalue $\ls\in \sigma(A(s))$ is isolated and
simple, such that
$$dist(\ls, \sigma(A(s))\backslash \{ \ls\})>d,$$
where $d>0$ is a constant independent of $s\in [0,1]$, and
$\ls$ is continuously differentiable in $s\in [0,1]$.

\item[(A5)] The projection onto $\ls$,
\begin{equation}
P_\ls:= \frac{1}{2\pi i}\oint_{\gamma_\ls} R(z,s) dz \; ,
\end{equation}
where $\gamma_\ls$ is a contour enclosing $\ls$ only, is twice
differentiable as a bounded operator.

\end{itemize}

Note that, since $\ls$ is simple, the resolvent of $A(s)$ in a neighborhood $\N$ of $\ls$
contained in a ball ${\mathbf B}(\ls,r)$ centered at $\ls$ with
radius $r<d$ is
\begin{equation}
\label{res} R(z,s)=\frac{P_\ls}{z-\ls} + R_{analytic} (z,s) \; ,
\end{equation}
where $R_{analytic}(z,s)$ is analytic in $\N .$ We recall some useful
properties of the resolvent and the spectral projection $P_\ls$; (see [Ka1]).

\begin{itemize}
\item[(i)] It follows by direct application of the contour
integration formula that
\begin{equation}
\label{sqrproj} (P_\ls)^2=P_\ls \; ,
\end{equation}
and hence
\begin{equation}
\label{dotP} 
P_\ls \dot{P}_\ls P_\ls=0 \; .
\end{equation}

\item[(ii)]
\begin{equation}
\label{projection} A(s)P_\ls = P_\ls A(s)=\ls P_\ls \;.
\end{equation}
{\it Proof.}
\begin{align*}
A(s)P_\ls &= \frac{1}{2\pi i}\oint_{\gamma_\ls}
(A(s)-z+z)(z-A(s))^{-1} dz \\
&= \frac{1}{2\pi i} \{ -\oint_{\gamma_\ls}dz
+\oint_{\gamma_\ls} (\frac{z P_\ls}{z-\ls} + zR_{analytic})dz \} \\
&= \ls P_\ls\; ,
\end{align*}
and similarly, $P_\ls A(s)=\ls P_\ls$.

\item[(iii)] It follows from (\ref{res}) and (A4) that, for $\eta\in {\mathbf C}$ and $\frac{d}{2}\le |\eta| < d,$
there exists a constant $C<\infty$, independent of $\eta,$
such that
\begin{equation}
\| R(\ls+\eta ,s) \| < C \; , \label{ResIneq}
\end{equation}
uniformly in $s\in [0,1]$. Moreover, since $(\ls+\eta ) \in \rho(A(s)),$ it follows by the spectral mapping theorem (see for example [Yo], Chp. VIII, section 7) and (A3) that $R(\ls+\eta ,s)$ is differentiable as a bounded operator.\footnote{We know that, for $z,\omega\in \rho(A),$ $$(z-A)^{-1}=(1+(z-\omega)(\omega-A)^{-1})^{-1}(\omega-A)^{-1}.$$ In particular, choose $z=\ls+\eta$ and $\omega=-1.$ Differentiability of $R(\ls+\eta)$ as a bounded operator follows from the latter identity and assumption (A3).}

\end{itemize}

We now discuss our general adiabatic theorem. Let $U_\tau(s,s')$ be the propagator satisfying
\begin{equation}
\label{rtevol2}
\partial_s U_\tau (s,s') = -\tau A(s) U_\tau (s,s')\; , U_\tau(s,s)=1 \; ,
\end{equation}
for $s\ge s'$.
Moreover, define the generator of the {\it adiabatic time
evolution},
\begin{equation}
A_a(s):= A(s)-\frac{1}{\tau}[\dot{P}_\ls,P_\ls] \; ,
\end{equation}
with the corresponding propagator $U_a(s,s')$ which satisfies
\begin{equation}
\label{ate2}
\partial_s U_a(s,s') = - \tau A_a(s) U_a (s,s') \; ; U_a(s,s)=1 \; ,
\end{equation}
for $s\ge s'$.

By Lemma 2.1 and (A1)-(A3) and (A5), both propagators
$U_\tau(s,s')$ and $U_a(s,s')$ exist and are unique, and $\|U_\tau (s,s')\| ,\|U_a(s,s')\| < C$ for $s\ge s'$, where $C$ is a finite constant independent of $s,s'\in [0,1]$.
We are in a position to state our adiabatic theorem.

\bigskip

\noindent {\bf Theorem 2.2 (A general adiabatic theorem)}

{\it Assume (A1)-(A5). Then the following holds.

\begin{itemize}
\item[(i)]
\begin{equation}
\label{intertwining}P_\ls U_a (s,0) =  U_a (s,0)P_\lambda (0) \;
,
\end{equation} 
for $s\ge 0$ (the intertwining property).
\item[(ii)] There is a finite constant $C$ such that
$$\sup_{s\in [0,1]}\| U_\tau(s,0)-U_a(s,0)\| \le \frac{C}{1+\tau},$$
for $\tau>0.$ In particular,
$$\sup_{s\in [0,1]} \| U_\tau (s,0)-U_a(s,0) \| =
O(\tau^{-1}),$$ as $\tau\rightarrow\infty.$
\end{itemize}}

\noindent {\it Proof.}
\begin{itemize}
\item[(i)] Equality holds trivially for $s=0$, since $U_a(s,s)=1$. Let 
\begin{equation}
h(s,s'):= U_a(s,s')P_\lsp U_a(s',0) \; ,
\end{equation}
for $0\le s' \le s.$


Using (\ref{projection}), (\ref{ate2}) , the definition of $A_a(s)$ and
the fact that $\dot{P}_\ls P_\ls + P_\ls \dot{P}_\ls =
\dot{P}_\ls$, it follows that
\begin{align*}
\partial_{s'}h(s,s') &= \partial_{s'}(U_a(s,s')P_\lsp U_a(s',0)) \\
&= \tau U_a(s,s') \{ A_a(s')P_\lsp
-P_\lsp A_a(s') \} U_a(s',0) \\&+ U_a (s,s') \dot{P}_\lsp U_a (s',0) \\
&= U_a(s,s') \{ -\dot{P}_\lsp P_\lsp \\&- P_\lsp \dot{P}_\lsp + \dot{P}_\lsp
\} U_a(s',0) \\
&= 0 \; .
\end{align*}
Therefore,
\begin{equation*}
h(s,s')\equiv h(s).
\end{equation*}
In particular,
\begin{equation*}
h(s,s)=h(s,0),
\end{equation*}
which implies claim (i).

\item[(ii)] Consider $\psi\in \D$, where the dense domain $\D$
appears in assumption (A2). We are interested in estimating the
norm of the difference $(U_\tau (s,0)-U_a (s,0))\psi$ as
$\tau\rightarrow \infty$. Using (\ref{rtevol2}), (\ref{ate2}) and the Duhamel formula,
\begin{align}
\label{diff} (U_\tau (s,0)-U_a (s,0))\psi  &=  - \int_0^s ds'
\partial_{s'} (U_\tau(s,s') U_a(s',0))\psi \\
&= \int_0^s  ds' (U_\tau(s,s')[\dot{P}_\lsp,P_\lsp]U_a(s',0))\psi \; .
\end{align}

Let\begin{equation}
X(s):= \frac{1}{2\pi i} \oint_{\gamma_\ls}dz R(z,s)\dot{P}_\ls R(z,s) \; ,
\end{equation}
where $\gamma_\ls$ is a contour of radius $d /2$ centered at $\ls,$ and where $d$ appears in (A4).
Then
\begin{eqnarray}
[X(s), A(s)] &=& \frac{1}{2\pi i}\oint_{\gamma_\ls}dz [z-A(s),R(z,s)\dot{P}_\lambda (s)R(z,s)] \nonumber \\
&=& \dot{P}_\ls P_\ls - P_\ls \dot{P}_\ls = [\dot{P}_\ls , P_\ls] \; .\label{commutator3}
\end{eqnarray}

Assumptions (A3),(A4) and the spectral mapping theorem imply that, for $z\in \gamma_\ls\subset\rho(A(s)), R(z,s)$ is differentiable as a bounded operator.
Together with (A5), this implies that,
\begin{eqnarray}
\|X(s)\| &<& C_1 \; , \label{est1}\\ 
\|\dot{X}(s)\| &<& C_2 \; , \label{est2}
\end{eqnarray} 
where $C_1$ and $C_2$ are finite constants independent of $s\in [0,1]$.
Moreover,
\begin{align*}
&U_\tau (s,s')[X(s'),A(s')]U_a(s',0) =\frac{1}{\tau}\{ -\partial_{s'}U_\tau(s,s')X(s')U_a(s',0) \\
&+U_\tau (s,s')(X(s')[\dot{P}_\lambda (s'),P_\lambda (s')])U_a(s',0)+U_\tau(s,s')\dot{X}(s')U_a(s',0)\} \; .
\end{align*}
Together with \fer{commutator3}, one may write the integrand in \fer{diff} as a total derivative plus a remainder term. Using the fact that $\D$ is dense in $\H$ and (A5),
\begin{equation}
\label{diff2} 
\| U_\tau (s,0) - U_a(s,0) \| \le 
\frac{1}{\tau }\sup_{s\in [0,1]}[C_1'\|X(s)\| + C_2' \| \dot{X}(s) \| ] \; ,
\end{equation}
where $C_i',i=1,2$ are finite constants independent of $s\in [0,1].$ 

Together with \fer{est1} and \fer{est2}, this implies
\begin{equation}
\sup_{s\in [0,1]}\|U_\tau (s,0) - U_a (s,0)\| \le \frac{C}{1+\tau} \; ,
\end{equation}
for $\tau>0,$ where $C$ is a finite positive constant. 
$\Box$

\bigskip

\end{itemize}

Next, we discuss a concrete model of a thermodynamic system to be studied subsequently.


\section{The Model}

As an example, we consider a two-level quantum system $\S$ coupled to $n$ reservoirs $,\R_1,\cdots ,\R_n,
n\ge 2,$ of free fermions in thermal equilibrium at inverse
temperatures $\beta_1, \cdots, \beta_n.$


\bigskip
\noindent{\bf The small system}
\bigskip

The kinematical algebra of $\S$ is $\O^\S=\M({\mathbf C}^2)$, the
algebra of complex $2\times 2$ matrices over the Hilbert space
$\H^\S={\mathbf C}^2$. Its Hamiltonian is given by $H^\S= \omega_0\sigma_3$, where $\sigma_i,i=1,2,3,$ are the Pauli matrices. When the system $\S$ is not coupled to the reservoirs, its dynamics in
the Heisenberg picture is given by
\begin{equation}
\a_\S^{t,s} (a) := e^{iH^\S (t-s)} a e^{-iH^\S (t-s)} \; ,
\end{equation}
for $a\in \O^\S.$

A physical state of the small system, $\omega^\S,$ is described by a density
matrix $\rho_\S .$ We assume that $\rho_\S>0,$ ie, $\omega^\S$ is faithful. The operator
$\kappa_\S=\rho_\S^{1/2}$ belongs to the space of Hilbert-Schmidt
operators, which is isomorphic to $\H^\S\otimes\H^\S.$ Two
commuting representations of $\O^\S$ on $\H^\S\otimes\H^\S$ are
given by
\begin{align}
& \pi_\S (a) := a\otimes\unit^\S \; , \\
& \pi_\S^\# (a) := \unit^\S \otimes C^\S a C^\S \; ,
\end{align}
where $C^\S$ is an antiunitary involution on $\H^\S$ 
corresponding to complex conjugation in the basis of the eigenvectors of $H^\S$; (see for example [BFS]).

The generator of the free dynamics on the Hilbert space
$\H^\S\otimes\H^\S$ is the standard Liouvillean
\begin{equation}
\L^\S = H^\S\otimes \unit^\S - \unit^\S\otimes H^\S \; .
\end{equation}

The spectrum of $\L^\S$ is $\sigma (\L^\S)=\{ -2\omega_0,0,2\omega_0 \}
,$ with double degeneracy at zero.

Let $\omega^\S$ be the initial state of the small system $\S,$
with corresponding vector $\Omega^\S\in \H^\S\otimes\H^\S.$ The
modular operator associated with $\omega^\S$ is $\Delta^\S =
\omega^\S\otimes \overline{\omega^\S}^{-1},$ and the modular
conjugation operator, $J^\S,$ is given by
$$J^\S (\phi\otimes\psi)=\overline{\psi}\otimes \overline{\phi},$$
for $\phi,\psi\in \H^\S.$
If $\omega^\S$ corresponds to the trace state, then $\Delta^\S=\unit^\S\otimes\unit^\S.$

\bigskip
\noindent{\bf The reservoirs}
\bigskip

Each thermal reservoir is formed of free fermions. It is
infinitely extended and {\it dispersive}. We assume that the
Hilbert space of a single fermion is $\h = L^2 ({\mathbf R}^+;\B),$
where $\B$ is an auxiliary Hilbert space, and $m(u)du$ is a measure on ${\mathbf R}^+$.
We also assume that the single-fermion Hamiltonian, $h,$ corresponds to the operator of
multiplication by $u\in {\mathbf R}^+ .$ For instance, for
reservoirs formed of nonrelativistic fermions in ${\mathbf R}^3,$
the auxiliary Hilbert space $\B$ is $L^2 (S^2, d\sigma),$ where
$S^2$ is the unit sphere in ${\mathbf R}^3,d\sigma$ is the uniform
measure on $S^2,$ and $u=|\vec{k}|^2,$ where $\vec{k}\in {\mathbf
R}^3$ is the particle's momentum. In the latter case, the measure
on ${\mathbf R}^+$ is choosen to be $m(u)du=\frac{1}{2}\sqrt{u}du.$

Let $b$ and $b^*$ be the annihilation-and creation operators on 
the Fermionic Fock space $\F(L^2({\mathbf R}^+; \B)).$ They satisfy the usual canonical anticommutation relation (CAR)
\begin{align}
&\{ b^\# (f), b^\# (g) \} = 0 \; , \\
&\{ b(f),b^* (g) \} = (f,g)\unit \; ,
\end{align}
where $b^\#$ stands for $b$ or $b^*,$ $f,g\in L^2 ({\mathbf R}^+ ;
\B),$ and $(\cdot , \cdot)$ denotes the scalar product in $L^2
({\mathbf R}^+ ; \B).$ Moreover, let $\Omega^\R$ denote the vacuum
state in $\F(L^2({\mathbf R^+};\B)) .$

The kinematical algebra, $\O^{\R_i},$ of the $i^{th}$ reservoir $\R_i
,i=1,\cdots ,n,$ is generated by $b_i^\#$ and the identity
$\unit^{\R_i}.$ The free dynamics of each reservoir (before the
systems are coupled) is given by
\begin{equation}
\a_{\R_i}^{t,s} (b_i^\# (f)) = b_i^\# (e^{i(t-s)u}f) \; ,
\end{equation}
for $i=1,\cdots ,n, f\in L^2 ({\mathbf R}^+ ; \B).$ 

The ($\a_{\R_i}, \beta_i$)-KMS state,
$\omega^{\R_i},$ of each reservoir $\R_i, i=1,\cdots , n,$ at
inverse temperature $\beta_i,$ is
the gauge invariant, quasi-free state uniquely determined by the
two-point function
\begin{equation}
\label{twopoint}
\omega^{\R_i} (b_i^*(f)b_i(f)) = (f, \rho_{\beta_i}(\cdot ) f) \; ,
\end{equation}
where $\rho_{\beta_i }(u) := \frac{1}{e^{\beta_i u}+1}.$ 

Next, we introduce $\F_i^{AW}:=\F^{\R_i}(L^2({\mathbf R}^+;\B))\otimes \F^{\R_i}(L^2({\mathbf R}^+;\B)),$ the GNS Hilbert space 
for the Araki-Wyss representation of each fermionic reservoir $\R_i$ associated with the state $\omega^{\R_i},$ [ArWy]. 
Denote by $\Omega^{\R_i}$ the vacuum state in $\F^{\R_i}(L^2({\mathbf R}^+;\B)),$ with 
$b_i\Omega^{\R_i}=0.$ The Araki-Wyss representation, $\pi_i,$ of the kinematical algebra $\O^{\R_i},
i=1,\cdots, n,$ on $\F^{AW}_i$  is given by 
\begin{eqnarray}
\pi_{i} (b_i (f)) &:=& b_i (\sqrt{1-\rho_{\beta_i}} \; f) \otimes \unit^{\R_i} + (-1)^{N_i}\otimes b_i^*(\sqrt{\rho_{\beta_i}} \; \overline{f}) \; ,\label{AW1} \\
\pi_{i}^\# (b_i (f)) &:=& b_i^* (\sqrt{\rho_{\beta_i}}f)(-1)^{N_i}\otimes (-1)^{N_i} +  \unit^{\R_i}\otimes (-1)^{N_i}b_i (\sqrt{1-\rho_{\beta_i }} \; \overline{f}) \; ,\nonumber
\end{eqnarray}
where $N_i=d\Gamma_i(1)$ is the particle number operator for
reservoir $\R_i.$ Furthermore, $\Omega^{\R_i}\otimes\Omega^{\R_i}\in\F_i^{AW}$ corresponds to the equilibrium 
KMS state $\omega^{\R_i}$ of reservoir $\R_i.$

The free dynamics on the GNS Hilbert space $\F_i^{AW}$ of each reservoir $\R_i$ is generated by the 
standard Liouvillean $\L^{\R_i}.$ The modular operator associated with $(\O^{\R_i},\omega^{\R_i})$ is given by
$$\Delta^{\R_i}=e^{-\beta_i \L^{\R_i}} \; ,$$
and the modular conjugation is given by
$$J^{\R_i}(\Psi\otimes\Phi)=(-1)^{N_i(N_i-1)/2}\overline{\Phi}\otimes (-1)^{N_i(N_i-1)/2}\overline{\Psi} ,$$
for $\Psi,\Phi\in \F_i^{AW};$ (see, for example, [BR]).


In order to apply the complex translation method developed in
[HP,JP1,2,3], we map $\F_i^{AW}=\F^{\R_i}(L^2 ({\mathbf R}^+ ; \B))\otimes
\F^{\R_i}(L^2 ({\mathbf R}^+ ; \B))$ to $\F^{\R_i}(L^2
({\mathbf R} ; \B))$ as done in [JP3]; (using the isomorphism between $L^2 ({\mathbf R}^+ ; \B)\oplus L^2 ({\mathbf R}^+ ; \B)$ and $L^2 ({\mathbf R}
; \B),$ the latter having measure $du$ on ${\mathbf R}).$ To every $f\in L^2 ({\mathbf R}^+ ; \B),$ we associate functions
$f_{\beta}, f_{\beta}^\# \in L^2 ({\mathbf R};\B),$ with measure $du$ on ${\mathbf R},$ by setting
\begin{equation}
f_{\beta} (u,\sigma) := \begin{cases}
\sqrt{m(u)}\sqrt{1-\rho_{\beta}(u)} f(u,\sigma) \; , & u\ge 0 \\
\sqrt{m(-u)}\sqrt{\rho_{\beta}(-u)} \; \overline{f}(-u,\sigma) \; , & u<0 \;
\end{cases} \; ,
\label{fglued}
\end{equation}
and
\begin{eqnarray}
f_{\beta}^{\#} (u,\sigma) &:=&
\begin{cases}
\sqrt{m(u)}i\sqrt{\rho_{\beta} (u)} f(u,\sigma),  u\ge 0 \\
\sqrt{m(-u)}i\sqrt{1-\rho_{\beta} (-u)} \; \overline{f}(-u,\sigma), u<0 \end{cases} \nonumber \\
&=& i\overline{f}_{\beta}(-u,\sigma) ,\label{fglued2}
\end{eqnarray}
where $m(u)du$ is the measure on ${\mathbf R}^+ ,$ see eq. \fer{AW1}.


Let $a_i$ and $a_i^*$ be the annihilation and creation operators on $\F^{\R_i}(L^2({\mathbf R}; \B))$. Then
\begin{align}
\pi_i(b_i(f)+b^*_i(f)) &\rightarrow a_i(f_{\beta_i})+a_i^*(f_{\beta_i}), \\
\pi_i^\#(b_i(f)+b_i^*(f)) &\rightarrow i (-1)^{N_i} [a_i(f^\#_{\beta_i})+a_i^*(f^\#_{\beta_i})] ;\\
\Omega^{\R_i}\otimes\Omega^{\R_i} &\rightarrow \tilde{\Omega}^{\R_i},
\end{align}
where $\tilde{\Omega}^{\R_i}$ is the vacuum state in 
$\F^{\R_i}(L^2({\mathbf R};\B)).$ \footnote{For a discussion of this map, see Theorem 3.3 in [JP3]; (see also the Appendix).} 

Moreover, the free Liouvillean on $\F^{\R_i}(L^2 ({\mathbf R};\B))$ for the reservoir $\R_i $ is mapped to
\begin{equation}
\L^{\R_i}=d\Gamma_i (u) \; ,
\end{equation}
where $u \in {\mathbf R}.$


\bigskip
\noindent{\bf The coupled system}
\bigskip

The kinematical algebra of the total system, $\S\vee\R_1\vee\cdots \vee
\R_n,$ is given by
\begin{equation}
\O=\O^\S\otimes \O^{\R_1}\otimes \cdots \otimes\O^{\R_n} \; ,
\end{equation}
and the Heisenberg-picture dynamics of the uncoupled system is given by
\begin{equation}
\a_0 = \a_\S \otimes \a_{\R_1} \otimes\cdots\otimes
\a_{\R_n} \; .
\end{equation}

The representation of $\O$ on 
$\H:= \H^\S\otimes\H^\S\otimes\F^{\R_1}(L^2 ({\mathbf R};\B))\otimes\cdots \otimes\F^{\R_n}(L^2 ({\mathbf R};\B)),$ 
determined by the initial state
\begin{equation}
\omega = \omega^\S\otimes \omega^{\R_1}\otimes\cdots\otimes \omega^{\R_n} 
\end{equation}
by the GNS construction, is given by
\begin{equation}
\pi=\pi_\S\otimes\pi_{1}\otimes\cdots\otimes\pi_{n} , \\
\end{equation}
and an anti-representation commuting with $\pi$ by
\begin{equation}
\pi^\# =\pi_\S^\#\otimes\pi_{1}^\# \otimes \cdots
\otimes\pi_{n}^\# \; .
\end{equation}
Moreover, let $\Omega:=\Omega^\S\otimes\tilde{\Omega}^{\R_1}\otimes\cdots\otimes\tilde{\Omega}^{\R_n}$ denote the vector in $\H$ corresponding to the state $\omega.$ Denote the double commutant of $\pi(\O)$ by $\M:=\pi(\O)'' ,$ which is the
smallest von Neumann algebra containing $\pi(\O)$.

For $a\in\O,$ we abbreviate $\pi (a)$ by $a$ whenever
there is no danger of confusion. The modular operator of the total system is
$$\Delta = \Delta^\S\otimes\Delta^{\R_1}\otimes\cdots\otimes\Delta^{\R_n} ,$$
and the modular conjugation is
$$J=J^\S\otimes J^{\R_1} \otimes\cdots\otimes J^{\R_n} .$$

According to Tomita-Takesaki theory,
$$J\M J=\M'\; , \Delta^{it} \M \Delta^{-it}=\M \; ,$$
for $t\in {\mathbf R};$ (see for example [BR]). Furthermore, for $a\in\M,$
\begin{equation}
J\Delta^{1/2}a\Omega=a^*\Omega .
\label{MCP}
\end{equation} 


The Liouvillean of the total uncoupled system is
given by
\begin{equation}
\L_0 = \L^\S +\sum_{i=1}^n \L^{\R_i}  \; .
\label{freeL}
\end{equation}
This defines a selfadjoint operator on $\H.$

The system $\S$ is coupled to the reservoirs $\R_1,\cdots ,\R_n,$ through an interaction $gV(t),$
where $V(t)\in\O$ is given by
\begin{equation}
V(t)=\sum_{i=1}^n \{ \sigma_1 \otimes [b_i (f_i (t)) +  b^*_i (f_i (t))] \} \; ,
\label{interactinghamiltonian}
\end{equation}
$\sigma_i, i=1,2,3,$ are the Pauli matrices,  
and $f_i \in L^2({\mathbf R}^+ ; \B), i=1,\cdots ,n,$
are the form factors.



The {\it standard} Liouvillean of the interacting system acting
on the GNS Hilbert space $\H$ is given by
\begin{equation}
\L_g (t) = \L_0 + gI(t)\; ,
\end{equation}
where the unperturbed Liouvillean is defined in \fer{freeL}, and the
interaction Liouvillean determined by the operator $V(t)$ is given by
\begin{align}
&I(t)=\{ V(t) - JV(t)J \} \nonumber\\
&=\sum_{i=1}^n \{ \sigma_1\otimes {\mathbf 1}^\S\otimes [a_i^*(f_{i,
\beta_i}(t))+ a_i(f_{i,\beta_i}(t))] \nonumber \\
&- i {\mathbf 1}^\S\otimes \sigma_1\otimes (-1)^{N_i} [a_i^*(f_{i,
\beta_i}^\#(t))+a_i(f_{i, \beta_i}^\# (t))] \} , 
\label{interaction}
\end{align}
where $a_i,a_i^*$ are the annihilation and creation operators on the fermionic Fock space $\F^{\R_i}(L^2({\mathbf R};\B)).$ Note that since the perturbation is bounded, the domain of $\L_g(t)$ is $\D(\L_g(t))=\D(\L_0).$ 

Let $\overline{U}_g$ be the propagator generated by the standard Liouvillean. It satisfies
\begin{equation}
\partial_t\overline{U}_g(t,t')=-i\L_g(t)\overline{U}_g(t,t')\; ; \overline{U}_g(t,t)=1 \; , 
\end{equation}
for $t\ge t'.$
The Heisenberg-picture evolution is given by 
\begin{equation}
\label{StandEvol}
\a_g^{t,t'}(a)=\overline{U}_g^*(t,t')a\overline{U}_g(t,t') \; ,
\end{equation}
for $a\in\O.$


Generally, the kernel of $\L_g(t), Ker \; \L_g,$ is expected to be empty when at least two of
the reservoirs have different temperatures.\footnote{This is consistent with the fact that the coupled system is not expected to possess the property of return to equilibrium if the reservoirs have different temperatures (or chemical potentials). One can verify that, indeed, this is the case when assumptions (B1) and (B2), below, are satisfied; (see [JP3,MMS1,2]).}. This
motivates introducing the so called C-Liouvillean, $L_g$ , which
generates an equivalent dynamics on a suitable Banach space contained in $\H$ (isomorphic to
$\O$) and which, {\it by construction}, has a non-trivial kernel.

Consider the Banach space
$$\C(\O,\Omega):=\{ a\Omega : a\in \O \} ,$$
with norm $\| a\Omega \|_{\infty}=\|a\|.$ Since $\Omega$ is
separating for $\O,$ the norm $\|a\Omega\|_\infty$ is well-defined, and since $\Omega$ is cyclic for $\O,$ 
$\C(\O,\Omega)$ is dense in $\H.$

Let $U_g(t,t')$ be the propagator given by
\begin{equation}
\a_g^{t,t'} (a)\Omega = U_g(t,t')a\Omega ,
\label{map}
\end{equation}
and \begin{equation}
U_g(t',t)\Omega = \Omega .
\end{equation}
Moreover, let $L_g(t)$ be its generator, ie, 
\begin{equation}
\partial_t U_g(t,t') = iU_g(t,t')L_g(t) \; {\mathrm with }\; U_g(t,t)=1 .
\label{propagator}\\
\end{equation}

Differentiating \fer{map} with respect to $t,$ setting $t=t',$ and using \fer{propagator}, \fer{StandEvol} and \fer{MCP}, one obtains
\begin{align*}
[(\L_0+gV(t))a-a(\L_0+gV(t))]\Omega &= [(\L_0+gV(t))a-(V(t)a^*)^*]\Omega \\
&= (\L_0+gV(t)-gJ\Delta^{1/2}V(t)\Delta^{-1/2}J)a\Omega \\
&\equiv L_g(t)a\Omega \; .
\end{align*} Hence, the C-Liouvillean is given by
\begin{equation}
\label{CL} L_g(t):= \L_0 + gV(t) -
gJ\Delta^{1/2}V(t)\Delta^{-1/2}J \; .
\end{equation}
Note that, {\it by construction},
$$L_g(t)\Omega=0,$$
for all $t\in {\mathbf R}.$

Next, we discuss the assumptions on the interaction. For
$\delta>0,$ we define the strips in the complex plane
$$I(\delta):= \{ z\in {\mathbf C} : | \Im z | < \delta \}$$
and
\begin{equation}
I^- (\delta):= \{ z\in {\mathbf C} : -\delta < \Im z < 0 \} .
\label{Iminus}
\end{equation}
Moreover, for every function $f\in L^2({\mathbf R^+};\B),$
we define a function $\tilde{f}$ by setting
\begin{equation}
\tilde{f}(u,\sigma):=
\begin{cases}
\sqrt{m(u)}f(u,\sigma) , u\ge 0  \\
\sqrt{m(|u|)} \; \overline{f}(|u|,\sigma) , u<0
\end{cases} ,\label{tildefi}
\end{equation}
where $m(u)du$ is the measure on ${\mathbf R}^+.$ Denote by $H^2(\delta , \B)$ the Hardy class of analytic
functions
$$h: I(\delta) \rightarrow \B , $$
with
$$\| h \|_{H^2 (\delta ,\B)} := \sup_{|\theta | < \delta} \int_{{\mathbf R}}\| h (u+i\theta )\|_\B^2 du < \infty . $$
We require the following basic assumptions on the interaction term.

\begin{itemize}

\item[(B1)] {\it Fermi Golden Rule.}

Assume that
\begin{equation}
\sum_{i=1}^n \| \tilde{f}_i (2\omega_0,t) \|_\B >0 \; ,
\end{equation}
for almost all $t\in {\mathbf R}$, which is another way of saying
that the small system is coupled to at least one reservoir, to second order in perturbation theory.

\item[(B2)] {\it Regularity of the form factors.}

Assume that $\exists \delta >0$, independent of $t$ and $i=1,\cdots,n,$ such that
\begin{equation}
e^{-\beta_i u/2}\tilde{f}_{i}(u,t) \in H^2(\delta,\B) \; ,
\end{equation}
the Hardy class of analytic functions. This implies that the mapping
\begin{equation}
{\mathbf R}\ni r\rightarrow \Delta^{ir} V(t)\Delta^{-ir}\in \M \; ,
\end{equation}
(where $\Delta = \Delta^\S\otimes\Delta^{\R_1}\otimes\cdots\otimes\Delta^{\R_n}$ is the modular operator of the coupled system, and $\M=\pi(\O)'',$) 
has an analytic continuation to the strip $I(1/2)=\{ z\in {\mathbf
C}: |\Im z|<1/2 \}$, which is bounded and continuous on its
closure, $\forall t\in {\mathbf R}$.

\item[(B3)] {\it Adiabatic evolution.}

The perturbation is constant for $t<0$, $V(t)\equiv V(0)$,
and then {\it slowly} changes over a time interval $\tau$ such
that $V^\tau (t)=V(s)$, where $s=t/\tau\in [0,1]$ is the rescaled
time. We also assume that $V(s)$ is twice differentiable in $s\in
[0,1]$ as a bounded operator, such that
\begin{equation}
{\mathbf R}\ni r\rightarrow \Delta^{ir}
\partial_s^j V(s)\Delta^{-ir}\in \M\; , j=0,1,2,
\end{equation}
has an analytic continuation to the strip $\{ z\in {\mathbf C}:
|\Im z|<1/2 \}$, which is bounded and continuous on its closure.
This follows if we assume that there exists $\delta >0$, independent of
$s$ and $i=1,\cdots,n,$ such that
\begin{equation}
e^{-\beta_i u/2}\partial_s^j \tilde{f}_{i}(u,s) \in H^2(\delta,\B)
\; ,
\end{equation}
the Hardy class of analytic functions, for $j=0,1,2.$ This
assumption is needed to prove an adiabatic theorem for states
close to NESS.\footnote{When the reservoirs are formed of nonrelativistic fermions in ${\mathbf R}^3,$ an example of a form factor satisfying assumptions (B1)-(B3) is given by  
$$f_i(u,s)=h_i(s)|u|^{1/4}e^{-|u|^2},$$ 
where $h_i(s)$ is twice differentiable in $s$.}

\end{itemize}

Let $\tilde{U}_g$ be the propagator generated by the adjoint of
the C-Liouvillean, ie,
\begin{eqnarray}
\partial_t \tilde{U}_g (t,t')&=& -iL_g^* (t)\tilde{U}_g(t,t') ,\\
\tilde{U}_g (t,t) &=& 1 .
\end{eqnarray}
Assumption (B2) implies that the perturbation is bounded, and
hence the domain of $L_g^\# ,$ where $L_g^\# $ stands for $L_g$ or
$L_g^*,$ is
$$\D(L_g^\#) = \D(\L_0) ,$$
and $U_g,\tilde{U}_g$ are bounded and strongly continuous in $t$ and $t'.$


\section{The C-Liouvillean and NESS}

In [JP3, MMS1,2], it is shown that, when the perturbation is
time-independent, and under reasonable regularity assumptions on the form factors, the state of the coupled system converges to a
nonequilibrium steady state (NESS) which is related to a zero-energy resonance of the adjoint of the C-Liouvillean. Here, we study the C-Liouvillean in the {\it time-dependent} case, and relate a zero-energy resonance to the {\it instantaneous} NESS. The statements made in this section have been proven in [JP3] (see also [JP1,2]) for the {\it time-independent} case. Extending those results to the {\it time-dependent} case is {\it straightforward}, since we study the spectrum of the Liouvillean at each {\it fixed} moment of time. However, a sketch of the proofs of all the statements made in this section is given in the Appendix to make the presentation self-contained.

We first study the spectrum of $L_g^*$ using complex spectral deformation
techniques as developed in [HP,JP1,2,3].

Let ${\mathbf u}_i$ be the unitary
transformation generating translations in energy for the $i^{th}$
reservoir, $i=1,\cdots,n.$ More precisely, for $f_i\in L^2({\mathbf R};\B),$
$${\mathbf u}_i(\theta) f_i (u) = f_i^{\theta}(u) = f_i (u+\theta)  .$$
Moreover, let
$$U_i(\theta):=\Gamma_i({\mathbf u}_i(\theta)) $$
denote the second quantization of ${\bf u}_i(\theta).$

Explicitly, $U_i(\theta)=e^{-i\theta A_i},$ where $A_i := i
d\Gamma_i (\partial_{u_i})$ is the second quantization of the
generator of energy translations for the $i^{th}$reservoir,
$i=1,\cdots ,n.$ We set
\begin{equation}
U(\theta) := \unit^\S\otimes\unit^\S\otimes U_1(\theta)
\otimes\cdots\otimes U_n(\theta) \; .
\end{equation}
Define
\begin{equation}
L^*_g(t, \theta) := U(\theta)L^*_g (t)U(-\theta) , \label{deformed}
\end{equation}
which is given by
\begin{equation}
L^*_g (t, \theta)= \L_0 +
N\theta + g\tilde{V}^{tot}(t,\theta) \; , \label{CL}
\end{equation}
$\L_0=\L^\S+\sum_i \L^{\R_i}$, $\L^{\R_i}=d\Gamma (u_i),
i=1,\cdots ,n$, $N=\sum_i N_i,$ the total particle number operator, and
\begin{align*}
\tilde{V}^{tot}(t,\theta)&= \sum_i \{ \sigma_1\otimes {\mathbf
1}^\S \otimes [a_i (f_{i,\beta_i}^{(\theta )}(t)) + a_i^* (f_{i,\beta_i}^{(\theta )}(t))] - i\unit^\S \otimes (\rho^\S)^{-1/2}\\&\sigma_1(\rho^{\S})^{1/2} \otimes (-1)^{N_i}[a_i(e^{\beta_i
u_i/2 }f_{i, \beta_i}^{\# (\theta )}(t))+ a_i^*(e^{-\beta_i
u_i/2}f_{i,\beta_i}^{\# (\theta )}(t))] \} \; .
\end{align*}
It follows from assuption (B2) that, for $\theta \in I(\delta),$
$\tilde{V}_g^{tot} (t,\theta)$ is a bounded operator. Hence $L^*_g(t,\theta)$ is well-defined and closed on the domain $\D:=\D(N)\cap\D(\L^{\R_1})\cap\cdots\cap \D(\L^{\R_n}).$ When the coupling $g=0$, the pure point spectrum of
$\L_0$ is $\sigma_{pp}(\L_0)=\{-2\omega_0,0,2\omega_0\}$,
with double degeneracy at 0, and the continuous spectrum of $\L_0$ is $\sigma_{cont}(\L_0)=\mathbf{R}$. Let $$\L_0(\theta):=\L_0 + N\theta .$$ We have the following two lemmas. 

\bigskip
\noindent {\bf Lemma 4.1}
{\it For $\theta\in {\mathbf C}$, the following holds.
\begin{itemize}

\item[(i)] For any $\psi\in \D$, one has
\begin{equation}
\| \L_0 (\theta ) \psi \|^2 = \| \L_0 (\Re\theta) \psi \|^2 + |
\Im \theta |^2 \| N \psi \|^2 \; .
\end{equation}

\item[(ii)] If $\Im \theta \ne 0$, then $\L_0 (\theta )$ is a normal operator satisfying
\begin{equation}
\L_0(\theta)^*=\L_0 (\overline{\theta}) \; ,
\end{equation}
and $\D(\L_0(\theta))=\D$.

\item[(iii)] The spectrum of $\L_0 (\theta )$ is
\begin{align}
\sigma_{cont} (\L_0 (\theta )) &= \{ n\theta + s : n\in {\mathbf N\backslash\{0\} } \;
{\mathrm and} \;  s\in {\mathbf R} \} ,\\
\sigma_{pp}(\L_0(\theta)) &= \{E_j : j=0,\cdots,3\} ,
\end{align}
where $E_{0,1}=0, E_2=-2\omega_0$ and $E_3=2\omega_0,$ (the eigenvalues of $\L^\S$).

\end{itemize}}

\bigskip

\bigskip
\noindent {\bf Lemma 4.2}

{\it Suppose assumptions (B1) and (B2) hold, and assume that $(g, \theta)\in {\mathbf
C}\times I^- (\delta ).$ Then, for each fixed time $t\in {\mathbf R},$ the following holds.

\begin{itemize}

\item[(i)] $\D(L^*_g (t,\theta ))=\D$ and $(L^*_g (t,\theta))^*=L_{\overline{g}}(t,\overline{\theta})$.

\item[(ii)] The map $(g, \theta) \rightarrow L^*_g(t,\theta)$ from ${\mathbf C}\times I^- (\delta)$
to the set of closed operators on $\H$ is an analytic family (of type A) in each variable separately; (see [Ka1], chapter V, section 3.2).

\item[(iii)] For $g\in {\mathbf R}$ finite and $\Im z$ large enough,
\begin{equation}
s-\lim_{\Im \theta\uparrow 0}(L^*_g(t,\theta)-z)^{-1}= (L^*_g(t,\Re
\theta)-z)^{-1}\; . \label{removeCD2}
\end{equation}

\end{itemize}}

\bigskip


We now apply degenerate perturbation theory, as developed in [HP], to compute the spectrum of $L_g^*(t,\theta).$ Using contour integration, one may define the projection onto the perturbed eigenstates of $L_g^*(t,\theta),$ for $\theta\in I^-(\delta).$ Let
\begin{equation}
P_{g} (t,\theta):= \oint_{\gamma} \frac{dz}{2\pi i}(z-L^*_g (t,\theta))^{-1} \; ,
\end{equation}
where $\gamma$ is a contour that encloses the
eigenvalues $E_j,j=0,\cdots,3,$ at a distance $d>0,$ such that, for sufficiently small $|g|$ (to be specified below) the contour also encloses $E_j(g,t),$ the isolated eigenvalues of $L_g^*(t,\theta).$ We let 
\begin{equation*}
P_0=\unit^{\S} \otimes \unit^\S\otimes |\tilde{\Omega}^{\R_n}\otimes\cdots\otimes\tilde{\Omega}^{\R_1}\rangle \langle \tilde{\Omega}^{\R_1}\otimes\cdots\otimes\tilde{\Omega}^{\R_n}| ,
\end{equation*}
where $\unit^\Sigma$ corresponds to the identity on $\H^\Sigma$ and $\tilde{\Omega}^{\R_i}$ corresponds to the vacuum state in $\F^{\R_i}(L^2({\mathbf R};\B)).$
Furthermore, we define 
\begin{equation}
T_{g}(t):= P_{0} P_{g} (t,\theta)P_{0}. \label{T}
\end{equation}

Consider the isomorphism
\begin{equation}
\label{quasiFL} 
S_{g}(t,\theta ):=
T_{g}^{-1/2}(t) P_{0} P_{g} (t,\theta): Ran(P_{g}(t,\theta))\rightarrow Ran(P_0)
\end{equation}
and its inverse\footnote{It follows from (\ref{DiffProj}), Theorem 4.3 (i) below, that $T_g(t)\rightarrow 1$ on $Ran (P_g(t,\theta))$ as $g\rightarrow 0,$ and hence $S_g(t,\theta)$ is a well-defined operator on $Ran (P_g(t,\theta)).$ By (\ref{T}), it has the right inverse $S_g^{-1}(t,\theta).$ Moreover, $dim Ran (P_g(t,\theta))=dim Ran (P_0)$ for $g$ small enough, and hence $S_g^{-1}(t,\theta)$ is the inverse of $S_g(t,\theta).$}
\begin{equation}
S_{g}^{-1}(t,\theta):= P_{g}(t,\theta)
P_{0} T_{g}^{-1/2}(t):
Ran(P_0) \rightarrow
Ran(P_{g}(t,\theta)) .
\end{equation}
We set
\begin{equation}
M_{g}(t):= P_{0} P_{g}
(t,\theta)L^*_g(t,\theta)P_{g}(t,\theta)P_{0}\;,
\end{equation}
and define the quasi-C-Liouvillean by
\begin{equation}
\label{qFL} 
\S_g(t) := S_g (t,\theta) P_g(t,\theta) L^*_g(t,\theta) P_g(t,\theta) S_g^{-1}(t,\theta) = T_g^{-1/2}(t)M_g(t)T_g^{-1/2}(t) .
\end{equation}
Let $k=min\{\delta,\frac{\pi}{\beta_1},\cdots,\frac{\pi}{\beta_n}\},$
where $\delta$ appears in assumption (B2), section 3, and $\beta_1,\cdots,\beta_n,$
are the inverse temperatures of the reservoirs $\R_1,\cdots,\R_n,$ respectively.
For $\theta \in I^-(k)$ (see \fer{Iminus}), we choose a parameter $\nu$ such that
\begin{equation}
\label{nu}
-k<\nu<0 \; \; {\mathrm and} \;  -k<\Im\theta<-\frac{k+|\nu|}{2}.
\end{equation}
We also choose a constant $g_1>0$ such that
\begin{equation}
g_1 C < (k-|\nu|)/2,
\label{g1}
\end{equation}
where
\begin{eqnarray}
C&:=&\sup_{\theta\in I(\delta),t\in {\mathbf R}}\| \tilde{V}^{tot}(t,\theta)\|
\label{C}\\
&\le & \sup_{t\in{\mathbf R}, z\in I(\delta)} \frac{\sqrt{2}}{2}\sum_i |1+e^{-\beta_i z}|^{-1/2} (3\| \tilde{f}_i(t)\|_{H^2(\delta,\B)}+ \|e^{-\beta_i u/2} \tilde{f}_i\|_{H^2(\delta,\B)}) ,\nonumber
\end{eqnarray}
which is finite due to assumption (B2).

\bigskip
\noindent {\bf Theorem 4.3}

{\it Suppose that assumptions (B1) and (B2) hold. Then, for $g_1>0$ satisfying \fer{g1}, $\theta \in I^-(k),$ $\nu$ satisfying \fer{nu}, and $t\in {\mathbf R}$ fixed, the following holds uniformly in $t$, ie, $g_1$ is independent of $t.$

\begin{itemize}

\item[(i)] If $|g| < g_1$, the essential spectrum of the operator $L^*_g(t,\theta)$ is contained in the half-plane ${\mathbf C}\backslash \Xi(\nu),$ where $\Xi (\nu):= \{ z\in {\mathbf C} : \Im z \ge \nu \}.$ Moreover, the discrete spectrum of $L_g^*(t,\theta)$ is independent of $\theta\in I^-(k)$. If $|g|< \frac{1}{2} g_1$, then the spectral projections $P_{g} (t,\theta),$
associated to the spectrum of $L^*_g (t,\theta)$ in the half-plane $\Xi (\nu),$ are analytic in $g$ and satisfy the estimate
\begin{equation}
\| P_{g}(t,\theta) - P_{0} \| < 1\; .\label{DiffProj}
\end{equation}

\item[(ii)]If $|g|<\frac{g_1}{2}$, then the
quasi-C-Liouvillean $\S_{g}(t)$ defined in
(\ref{qFL}) depends analytically on $g$, and has a Taylor
expansion
\begin{equation}
\label{FLTaylor2} 
\S_{g}(t)=\L^\S +
\sum_{j=1}^{\infty} g^{2j} \S^{(2j)}(t).
\end{equation}
The first non-trivial coefficient in (\ref{FLTaylor2}) is
\begin{align*}
\S^{(2)}(t)& = \frac{1}{2} \oint_\gamma \frac{dz}{2\pi
i}(\xi(z,t)(z-\L^\S)^{-1}+(z-\L^\S)^{-1}\xi(z,t))\; ,
\label{2orderqFL}
\end{align*}
where $\xi(z,t):=P_{0}
\tilde{V}^{tot}(t,\theta)(z-\L_0(\theta))^{-1}\tilde{V}^{tot}(t,\theta)P_{0}.$

\end{itemize}}

\bigskip


In fact, one may apply second order perturbation theory to calculate the perturbed eigenvalues of $L_g^*(t, \theta).$ To second order in the coupling $g$, 
\begin{align*}
E_0(g,t)&=0 \; ,\\
E_1(g,t)&=-i \pi g^2 \sum_i \| \tilde{f}_i (2\omega_0,t) \|^2_\B
+O(g^4)\; ,
\end{align*}
and
\begin{align*}
E_{2,3}(g,t)=&\mp (2\omega_0 - \frac{1}{2}g^2 \PV\int_{\mathbf R} du \frac{1}{2\omega_0-u}
\sum_{i} \| \tilde{f}_i(u,t)\|^2_{\B})\\
 &-i\frac{\pi}{2} g^2 \sum_{i} \|\tilde{f}_i(2\omega_0,t)\|^2_\B + O(g^4) \; ,
\end{align*}
where $\PV$ denotes the Cauchy principal value (see the Appendix).


The following corollary follows for the case of {\it time-independent} interactions; (see [JP3,MMS1,2]). 

Define
\begin{equation}
D:= {\mathbf 1}^\S\otimes {\mathbf 1}^\S \otimes e^{-k
\tilde{A}_{\R_1}}\otimes\cdots\otimes e^{-k\tilde{A}_{\R_n}} \; ,\label{D}
\end{equation}
where $\tilde{A}_{\R_i}:=d\Gamma(\sqrt{p_i^2+1})$, and
$p_i:=i\partial_{u_i}$ is the generator of energy translations for
$\R_i$, $i=1,\cdots,n.$ Note that $D$ is a positive bounded operator on $\H$ such that $Ran (D)$ is dense in $\H$ and $D\Omega=\Omega$. This operator will act as a regulator which is used to apply complex deformation techniques. Let $\a_g^t\equiv \a_g^{t,0}.$

\bigskip

\noindent {\bf Corollary 4.4 (NESS)}

{\it Suppose assumptions (B1) and (B2) hold, and that the perturbation
$V(t)\equiv V$ is {\it time-independent}. Then there exists
$g_1>0$ such that, for $0<|g|<g_1$ and $a\Omega\in \D(D^{-1})$, the
following limit exists,
\begin{equation}
\lim_{t\rightarrow\infty} \langle \Omega ,\a^t_g (a) \Omega
\rangle = \langle \Omega_g, D^{-1} a \Omega \rangle \; ,
\end{equation}
where $\Omega_g$ corresponds to the zero-energy resonance of
$L_g^*$, and $\a_g^t$ is the perturbed dynamics. For $a\in \O^{test},$ a dense subset of $\O$ (that will be specified below), this limit is exponentially fast, with relaxation time $\tau_R=O(g^{-2}).$}\footnote{In fact, by assuming additional analyticity of the interacting Hamiltonian, one may show that this result holds for any initial state normal to $\omega$; see [JP3,MMS1,2].}

\bigskip

Moreover, [JP3,MMS1,2] prove strict positivity of entropy production in the latter case, which is consistent with Clausius' formulation of the second law of thermodynamics. See [FMUe] for another proof using scattering theory of the convergence to a NESS and strict positivity of entropy production when two free fermionic reservoirs at different temperatures or chemical potentials are coupled.


\section{Quasi-static evolution of NESS}

In this section, we apply Theorem 2.2, section 2, to investigate the quasi-static evolution of NESS of the model system introduced in section 3.

Together with assumption (B1), we assume (B3), ie, 
$V^{\tau}(t)=V(s),$ where $s\in[0,1]$ is the rescaled time with
sufficient smoothness properties of the interaction. From
Theorem 4.3, section 4, we know the spectrum of the deformed adjoint of
the C-Liouvillean, $L^*_g(t,\theta)=U(\theta)L^*_g(t)U(-\theta),$
for $\theta\in I^-(k),$ where
$k=\min (\delta,\frac{\pi}{\beta_1},\cdots,\frac{\pi}{\beta_n})$,
and $\delta$ appears in assumption (B3). Let $\gamma_0$ be a
contour enclosing only the zero eigenvalue of $L^*_g(s,\theta)$,
for all $s\in [0,1]$, and
\begin{equation}
\label{dproj1} 
P^0_g(s,\theta):= \oint_{\gamma_0}\frac{dz}{2\pi i}
(z-L^*_g(s,\theta))^{-1},
\end{equation}
the spectral projection onto the state corresponding to the zero eigenvalue of $L^*_g(s,\theta).$ Moreover, let $\h^{test}=
\D(e^{k\sqrt{p^2+1}})$, and $\O^{\R, test}$ be the algebra generated by $b^\#(f),f\in\h^{test},$ and $\unit^{\R}.$ Note that $\O^{\R, test}$
is dense in $\O^{\R}$. Define 
\begin{equation}
\O^{test}:= \O^\S\otimes\O^{\R_1 ,test}\otimes\cdots \O^{\R_n, test} \; ,
\end{equation}
which is dense in $\O,$ and 
$${\mathcal C}:= \{a\Omega : a\in \O^{test} \} \equiv \D(D^{-1}),$$
where $D$ is the positive operator as defined in (\ref{D}), section 4.
We make the following additional assumption.

\begin{itemize}

\item[(B4)] The perturbation Hamiltonian $V(s)\in \O^{test},$ for $s\in [0,1]$. 
\end{itemize}
In order to characterize the quasi-static evolution of nonequilibrium steady states, we introduce the new notion of an {\it instantaneous} NESS. Define an {\it instantaneous} NESS vector to be
\begin{equation}
\Omega_g(s):= DU(-\theta)P^0_g(s,\theta)U(\theta)D \Omega \; . \label{OmegaG}
\end{equation}
Note that $\Omega_g$ from Corollary 4.4, section 4, has the same form as (\ref{OmegaG}).

It is important to note that introducing the operator $D$ is needed to remove the complex deformation. 

We have the following Theorem, which effectively says that if a
system, which is initially in a NESS, is perturbed slowly over a
time scale $\tau\gg \tau_R$, where $\tau_R$ is some generic time
scale ($\tau_R=\max_{s\in [0,1]}\tau_{R(s)}$, and $\tau_{R(s)}$
is the relaxation time to a NESS, see proof of Corollary 4.4 in the Appendix), then the
real state of the system is infinitesimally close to the {\it
instantaneous} NESS, and the difference of the two states is
bounded from above by a term of order $O(\tau^{-1})$.

\vspace{0.5cm}

\noindent {\bf Theorem 5.1 (Adiabatic Theorem for NESS)}

{\it Suppose assumptions (B1), (B3) and (B4) hold. Then there exists $g_1>0$,
independent of $s\in [0,1]$, such that, for $a\in \O^{test}, s\in [0,1]$, 
and $0<|g|<g_1$, the following estimate
holds
\begin{equation}
\sup_{s\in [0,1]} | \langle \Omega_g(0), D^{-1} \a_g^{\tau s}(a)
\Omega\rangle - \langle \Omega_g(s), D^{-1} a \Omega\rangle | =
O(\tau^{-1}) \; ,
\end{equation}
as $\tau\rightarrow\infty$.}

{\it Proof.} Note that assumption (B3) implies assumption (B2), and hence the results of Theorem 4.3 about the spectrum of $L^*_g(t,\theta),$ for $\theta\in I^-(k)$ and {\it fixed} $t\in {\mathbf R},$ hold. The proof is now reduced to showing that the
assumptions of Theorem 2.2 are satisfied. Choose $\theta\in
I^-(k)$. It follows from assumption (B3) and Lemma A.1 in the Appendix, that the deformed C-Liouvillean $L^*_g(s,\theta)$
with common dense domain $\D=\D(\L_0)\cap \D(N) $ generates the propagator
$\tilde{U}^{(\tau)}_g(s,s',\theta), s'\le s,$ which is given by
\begin{equation}
\partial_s \tilde{U}^{(\tau)}_g(s,s',\theta) = -i \tau L^*_g(s,\theta)
\tilde{U}^{(\tau )}_g(s,s',\theta) \; ,\; {\rm for} \; s'\le s ; \tilde{U}_g^{(\tau)}(s,s,\theta)= 1 .
\end{equation}

This implies that (A1) and (A2) are satisfied. Furthermore,
(A3) follows from the second resolvent identity
\begin{equation}
\label{dr} (L^*_g (s, \theta)-z)^{-1} = (\L_0 (\theta
)-z)^{-1}(1+g \tilde{V}^{tot}(s,\theta)(\L_0 (\theta
)-z)^{-1})^{-1} \; ,
\end{equation}
and the results of Theorem 4.3, section 4. We also know that
zero is an isolated simple eigenvalue of $L^*_g(s,\theta )$ such
that $dist(0, \sigma(L^*_g(s,\theta))\backslash \{ 0 \})>d $,
where $d >0$ is a constant independent of $s\in [0,1]$. This
implies that assumption (A4) holds. Again using the resolvent
equation (\ref{dr}) and assumption (B3), $P^0_g(s,\theta)$
defined in (\ref{dproj1}) is twice differentiable as a bounded
operator for all $s\in [0,1]$, which imply (A5). Let $\tilde{U}^{(\tau
)}_a(s,s',\theta)$ (with domain $\D$) be the propagator of the {\it deformed} adiabatic
evolution given by
\begin{equation}
\label{dae}
\partial_s \tilde{U}^{(\tau )}_a(s,s',\theta)=
-i\tau L^*_a(s,\theta)\tilde{U}^{(\tau )}_a(s,s',\theta)\; {\rm for} \; s'\le s \; ;
\tilde{U}^{(\tau )}_a(s,s,\theta)=1 \; ,
\end{equation}
and
\begin{equation}
L^*_a(s,\theta) = L^*_g(s,\theta)
+\frac{i}{\tau}[\dot{P}_g(s,\theta),P_g(s,\theta)] \; .
\end{equation}
(Here, the $\dot{()}$ stands for differentiation with respect to
$s$.) Since (A1)-(A5) are satisfied, the results of Theorem 2.2 hold.
\begin{equation}
P^0_g(s,\theta) \tilde{U}^{(\tau)}_a
(s,0,\theta) = \tilde{U}^{(\tau)}_a(s,0,\theta)P^0_g (0,\theta) \; ,
\end{equation}
and
\begin{equation}
\sup_{s\in [0,1]}\| \tilde{U}^{(\tau
)}_g(s,0,\theta)-\tilde{U}^{(\tau )}_a(s,0,\theta) \| = O(\tau^{-1})
\; ,
\end{equation}
as $\tau\rightarrow\infty.$ 

For $h$ the single particle Hamiltonian of the free fermions, $e^{ih t}$ leaves $D(e^{k\sqrt{p^2+1}})$ invariant. Therefore, for $a\in \O^{test}$, $\a_0^t (a)\in \O^{test}$, where $\a_0^t$ corresponds to the free time evolution. Moreover, together with assumption (B4) and the boundedness of $V$, this implies (using a Dyson series expansion) that $\a_g^{\tau s}(a)\in \O^{test}.$ 

Now, applying the time evolution on $C(\O,\Omega)$, and remembering
that $D\Omega=\Omega$, $U(\theta)\Omega=\Omega$, the fact that
$U(\theta)$ and $D$ commute,  and the definition of the
instantaneous NESS, it follows that
\begin{equation}
\langle \Omega_g (0) ,D^{-1}\a_g^{\tau s}(a)\Omega\rangle 
=\langle \tilde{U}_g^{(\tau )}(s,0,\theta) P^0_g(0,\theta) \Omega, a(\overline{\theta})\Omega \rangle \; .
\end{equation}
Using the results of Theorem 2.2, it follows that 
\begin{align*}
&\langle \tilde{U}_g^{(\tau )}(s,0,\theta) P^0_g(0,\theta) \Omega, a(\overline{\theta})\Omega \rangle \\
&= \langle \tilde{U}_a^{(\tau )}(s,0,\theta) P^0_g(0,\theta) \Omega, a(\overline{\theta})\Omega \rangle + O(\tau^{-1}) \\
&= \langle P^0_g(s,\theta) \tilde{U}_a^{(\tau )}(s,0,\theta) \Omega, a(\overline{\theta})\Omega \rangle + O(\tau^{-1}) \\
&= \langle P^0_g(s,\theta) \tilde{U}_g^{(\tau )}(s,0,\theta) \Omega, a(\overline{\theta})\Omega \rangle + O(\tau^{-1}) \; .
\end{align*}
The fact that $(\tilde{U}_g^{(\tau )}(s,0,\theta))^*\Omega=\Omega$ implies 
\begin{align*}
D P^0_g(s,\theta) \tilde{U}_g^{(\tau )}(s,0,\theta) &= | \Omega_g(s,\theta) \rangle\langle \Omega | \tilde{U}_g^{(\tau )}(s,0,\theta) \\
&= | \Omega_g(s,\theta) \rangle\langle (\tilde{U}_g^{(\tau )}(s,0,\theta))^*\Omega | \\
&= | \Omega_g(s,\theta) \rangle\langle \Omega | = D P^0_g(s,\theta) \; .
\end{align*}
It follows that 
\begin{equation*}
\langle \Omega_g (0) ,D^{-1}\a_g^{(\tau s)}(a)\Omega\rangle 
= \langle \Omega_g(s) , D^{-1} a \Omega \rangle +O(\tau^{-1})\;,
\end{equation*}
for large $\tau$.
$\Box$

\bigskip

\noindent {\it Remarks.}

\begin{itemize}

\item[(1)] {\it Positivity of entropy production.} If the interaction Hamiltonian $gV(t)$ is time-periodic with period $\tau,$ ie, $V(t+\tau)=V(t),$ it is shown in [A-SF3] that the final state of the coupled system (introduced in section 3) converges to a time periodic state after very many periods. It is also shown that entropy production per cycle is strictly positive  (Theorem 6.3 in [A-SF3]). The infinite period limit, $\tau\rightarrow\infty,$ is equivalent to the quasi-static limit. Hence, entropy production in the quasi-static evolution of NESS of the model considered in this paper is strictly positive.

\item[(2)] {\it An example of a reversible isothermal process.} As a second application of Theorem 2.2 in quantum statistical mechanics, one may consider a concrete example of an isothermal process of a small system coupled to a {\it single} fermionic reservoir, and calculate an explicit rate of convergence ($O(\tau^{-1})$) between the instantaneous equilibrium state and the true state of the
system in the quasi-static limit $\tau\rightarrow\infty$ (see [A-SF1]). Under suitable assumptions on the form factors, one may show that  there exists a constant $g_1>0$ such
that, for $a$ in a dense subset of $\O$ and $0<|g|<g_1$, the following
estimate holds
\begin{equation}
| \rho_{\tau s}(a) - \omega_{\tau s}^\beta (a) | = O(\tau^{-1})
\; ,
\end{equation}
as $\tau\rightarrow\infty$, where $\rho_{\tau s}$ is the true
state of the  system at time $t=\tau s$, and $\omega_{\tau
s}^\beta$ is the instantaneous equilibrium state, which corresponds to the zero eigenvalue of the time-dependent {\it standard} Liouvillean.

\end{itemize}


\section{Appendix }


\bigskip
\noindent {\bf Existence of the {\it deformed} time evolution}

\bigskip

Choose $\theta\in I^-(\delta)$, where $\delta$ appears in
assumption (B2), section 3. The {\it deformed} time evolution is
given by the propagator $\tilde{U}_g(t,t',\theta)$ which satisfies
\begin{equation*}
\label{dte}
\partial_t \tilde{U}_g(t,t',\theta) = -i L_g^*(t,\theta)
\tilde{U}_g(t,t',\theta)\; ,\; \tilde{U}_g(t,t,\theta)=1 \; .
\end{equation*}

The following Lemma guarantees the existence of $\tilde{U}_g(t,t',\theta)$. Let
\begin{equation*}
\D:= \D(\L_0)\cap\D(N),
\end{equation*}
and denote by
\begin{eqnarray*}
\label{supInt} C&:=& \sup_{t\in\mathbf{R}}\sup_{\theta\in
I^-(\delta)} \| \tilde{V}^{tot}(t,\theta) \| \\
&\le& \frac{\sqrt{2}}{2}\sup_{t\in {\mathbf R}, z\in I(\delta)}\sum_i |1+e^{-\beta_i z}|^{-1/2}(3\| \tilde{f}_i(t)\|_{H^2(\delta, \B)}+ \| e^{-\beta_i u_i/2}\tilde{f}_i(t)\|_{H^2(\delta,\B)}) <\infty 
\end{eqnarray*}
due to assumption (B2), section 3.

\bigskip

\noindent {\bf Lemma A.1}

{\it Assume (B2), choose  $\theta\in I^-(\delta) \cup \mathbf{R}$
and $|g|< g_1 $, and fix $t \in \mathbf{R}$. Then

\begin{itemize}
\item[(i)] $L_g^*(t,\theta )$ with domain $\D$ generates a contraction semi-group
$e^{-i\sigma L_g^*(t, \theta)},\sigma\ge 0$ on $\H.$

\item[(ii)] For $\psi\in\D,$ $e^{-i\sigma
L_g^*(t,\theta)}\psi$ is analytic in $\theta\in I^-(\delta)$. For
$\theta'\in\mathbf{R}$ and $\theta\in I^-(\delta)\cup\mathbf{R}$,
\begin{equation*}
\label{twicedeformation} U(\theta') e^{-i\sigma
L_g^*(t,\theta)}U(-\theta')= e^{-i\sigma L_g^*(t,\theta + \theta')} \; .
\end{equation*}

\item[(iii)] $\tilde{U}_g(t,t',\theta)\tilde{U}_g(t',t'',\theta)=\tilde{U}_g(t,t'',\theta)$
for $t\ge t'\ge t''$.

\item[(iv)] $\tilde{U}_g(t,t',\theta)\D\subset\D$, and for
$\psi\in\D$, $\tilde{U}_g(t,t',\theta)\psi$ is differentiable in
$t$ and $t'$ such that
\begin{align*}
\partial_t \tilde{U}_g(t,t',\theta) \psi&= -i
L_g^*(t,\theta)\tilde{U}_g(t,t',\theta)\psi\; , \\
\partial_{t'}\tilde{U}_g(t,t',\theta)\psi &=
i\tilde{U}_g(t,t',\theta)L_g^*(t',\theta) \psi\; .
\end{align*}

\item[(v)] For $\theta'\in\mathbf{R}$,
\begin{equation*}
U(\theta')\tilde{U}_g(t,t',\theta) U(-\theta') = \tilde{U}_g(t,t',\theta + \theta') \; .
\end{equation*}
Moreover, $\tilde{U}_g(t,t',\theta)$ is analytic in $\theta\in
I^-(\delta)$.

\end{itemize}}

{\it Proof.} Claim (i) follows from Phillip's Theorem for the
perturbation of semigroups (see [Ka1] chapter IX). Claim (ii) follows
from assumption (B2), the resolvent identity
\begin{equation*}
(L_g^*(t,\theta)-z)^{-1} = (\L_0(\theta)-z)^{-1}
(1+\tilde{V}^{tot}(t,\theta)(\L_0(\theta)-z)^{-1})^{-1} \; ,
\end{equation*}
\begin{equation*}
U(\theta')L_g^*(t,\theta)U(-\theta')=L_g^*(t,\theta + \theta') \; ,
\end{equation*}
and the fact that
\begin{equation*}
e^{-i\sigma L_g^*(t,\theta)}=\frac{1}{2\pi i}\int_{\Gamma} e^{-\sigma
z} (iL_g^*(t,\theta)-z)^{-1}dz \; ,
\end{equation*}
where $\Gamma$ is a contour encircling the spectrum of $L_g^*(t, \theta)$.

Claims (iii) and (iv) are consequences of Kato's
Theorem [Ka2], to which we refer the reader. Without loss of
generality, rescale time such that $t=\tau s, s\in [0,1]$, and let
$L_g^{*n}(s\tau,\theta)=L_g^*(\tau\frac{k}{n},\theta)$ for $n\in
{\mathbf N}\backslash \{ 0 \}$ and $s\in[\frac{k}{n},\frac{k+1}{n}],
k=0,\cdots,n-1$. Moreover, define $\tilde{U}_g^{n}(\tau s,\tau
s',\theta):=e^{-i\tau(s-s')L_g^{*n}(\tau\frac{k}{n},\theta)}$ if
$\frac{k}{n}\le s'\le s \le \frac{k+1}{n}$, and $\tilde{U}_g^n(\tau s,\tau s',\theta)=\tilde{U}_g^n(\tau s,\tau
s'',\theta)\tilde{U}_g^n(\tau s'',\tau s',\theta)$ if $0\le s'\le
s'' \le s \le 1$. It follows from (ii) for
$\theta'\in\mathbf{R}$, that
\begin{equation*}
U(\theta')\tilde{U}_g^n (\tau s, \tau s' ,\theta) U(-\theta') =
\tilde{U}_g^n(\tau s, \tau s', \theta +\theta') \; ,
\end{equation*}
and that $\tilde{U}_g^n(\tau s, \tau s', \theta)$ is analytic in $\theta\in
I^-(\delta)$, where $\delta$ appears in (B2). Claim (v) follows by taking the $n\rightarrow\infty$ limit (in norm). 
$\Box$

\bigskip
\bigskip

\noindent{\bf Glued Hilbert space representation}
\bigskip

We want to show that
\begin{equation*}
\F(L^2({\mathbf R}^+;\B))\otimes \F(L^2({\mathbf R}^+;\B)) \cong \F(L^2 ({\mathbf R};\B)) \; .
\end{equation*} 
Let $\Omega$ be the vacuum state in the fermionic Fock space $\F(L^2({\mathbf R}^+;\B)).$ For fermionic creation/annihilation operators on $\F(L^2({\mathbf R}^+; \B)),$ 
$$b^\# (f):= \int m(u)dud\sigma f(u,\sigma)b^\# (u,\sigma) \; , f \in L^2({\mathbf R}^+; \B),$$
define the creation/annihilation operators on $\F(L^2({\mathbf R}^+;\B))\otimes \F(L^2({\mathbf R}^+;\B))$ as
\begin{align*}
b_l^\# (f) := b^\# (f)\otimes \unit \; ; \\
b_r^\# (f) := (-1)^{N} \otimes b^\# (\overline{f}) \; ,
\end{align*}
where $\overline{\cdot}$ corresponds to complex conjugation. Note that $b_l$ and $b_r$ anti-commute. Let $\tilde{a}$ and $\tilde{a}^*$ be the annihilation and creation operators on the fermionic Fock space $\F(L^2({\mathbf R}^+;\B)\oplus L^2({\mathbf R}^+;\B)),$ such that they satisfy the usual CAR, and let $\tilde{\Omega}$ be the vacuum state in $\F(L^2({\mathbf R}^+;\B)\oplus L^2({\mathbf R}^+;\B)).$
An isomorphism between $\F(L^2({\mathbf R}^+;\B))\otimes \F(L^2({\mathbf R}^+;\B))$
and $\F(L^2({\mathbf R}^+;\B)\oplus L^2({\mathbf R}^+;\B))$ follows by the identification
\begin{align*}
b_l^\# (f)&\cong \tilde{a}^\# ((f,0)) ,\\
b_r^\# (g) &\cong \tilde{a}^\# ((0,g)) \; ,\\
\Omega\otimes\Omega&\cong \tilde{\Omega}.
\end{align*}

Now we claim that $\F(L^2({\mathbf R}^+;\B)\oplus L^2({\mathbf R}^+;\B))$ is isomorphic to
$\F(L^2 ({\mathbf R};\B))$. Consider the mapping
\begin{equation*}
j : L^2({\mathbf R}^+;\B)\oplus L^2({\mathbf R}^+;\B) \ni (f,g)\rightarrow h \in L^2 ({\mathbf R};\B) \; ,
\end{equation*}
such that
\begin{equation*}
h(u,\sigma) :=
\begin{cases}
\sqrt{m(u)} f(u,\sigma) \; , u\ge 0 \\
\sqrt{m(|u|)} g(|u|,\sigma) \; , u < 0
\end{cases} \; .
\end{equation*}
This mapping is an isometry, since
\begin{align*}
\| h \|^2_{L^2 ({\mathbf R};\B)} &= \| (f,g) \|^2_{ L^2({\mathbf R}^+;\B) \oplus L^2({\mathbf R}^+;\B) } \\
&= \int_{{\mathbf R}^+;\B} du d\sigma m(u) |f(u,\sigma)|^2 + \int_{{\mathbf R}^+;\B} du d\sigma m(u) |g(u,\sigma)|^2 \\
&= \| f \|^2_{L^2({\mathbf R}^+;\B)} + \| g \|^2_{L^2({\mathbf R}^+;\B)} \; .
\end{align*}
Moreover, the mapping $j$ is an isomorphism, since, for given
$h\in L^2 ({\mathbf R};\B)$, there exists a mapping
$j^{-1}: h\rightarrow (f,g)\in L^2({\mathbf R}^+;\B)\oplus L^2({\mathbf R}^+;\B)$, such that
\begin{align*}
f(u,\sigma)&:= \frac{1}{\sqrt{m(u)}}h(u,\sigma) , u>0 \; , \\
g(u,\sigma)&:= \frac{1}{\sqrt{m(|u|)}}h(|u|,\sigma) , u<0 \; .
\end{align*}

\bigskip
\noindent{\bf Proof of statements in Section 4} \footnote{Although the results in this subsection are a very simple extension of those proven in [JP1,2,3] to the time-dependent case, they are sketched here so that the presentation is self-contained. The reader can refer to those references for additional details.}

\bigskip
\noindent {\it Proof of Lemma 4.1}

$\L_0 (\theta)$ restricted to the $N=n\unit$ sector is
\begin{equation}
\L_0^{(n)}(\theta )=\L^\S + s_1 + \cdots + s_n + n\theta\; ,\label{sector}
\end{equation}
where $s_1,\cdots,s_n$ are interpreted as one-particle multiplication operators.
For $\Im\theta\ne 0,$ it also follows from \fer{sector}
that 
$$\D = \{ \psi = \{ \psi^{(n)} \} : \psi^{(n)} \in
\D(\L_0^{(n)}(\theta)) \; {\mathrm and} \; \sum_n \|
\L_0^{(n)}(\theta) \psi^{(n)}\|^2 < \infty \},$$ and hence 
$\L_0 (\theta )$ is a closed normal operator on $\D$. Claims (ii)
and (iii) follow from the corresponding statements on the sector $N=n\unit.$ $\Box$

\bigskip


\noindent {\it Proof of Lemma 4.2.}

 The first claim (i) follows from the fact that
$g\tilde{V}^{tot}(t, \theta)$ is bounded for $\theta\in I(\delta)$ due to assumption (B2) and the fact that the reservoirs are fermionic. It also follows from assumption (B2) that $(g, \theta)\rightarrow L^*_g(t, \theta)$ is analytic in $\theta\in I^-(\delta).$
Analyticity in $g$ is obvious from \fer{deformed}. Assume that $\Re \theta = 0$. It follows from assumption (B2) that the resolvent formula
\begin{equation}
\label{resolvent} (L^*_g (t,\theta)-z)^{-1} = (\L_0 (t,\theta
)-z)^{-1}(1+ g\tilde{V}^{tot}(t,\theta )(\L_0 (\theta )-z)^{-1})^{-1} \; ,
\end{equation}
holds for small $g$, as long as $z$ belongs to the half-plane $\{ z\in
{\mathbf C}: 0< c< \Im z \}.$  Since $(\L_0(t,\theta)-z)^{-1}$ is uniformly bounded as $\Im\theta\uparrow 0$ for $g\in {\mathbf R}$ and $\Im z$ large enough, and $\tilde{V}^{tot}(t,\theta)$ is bounded and analytic in $\theta$, claim (iii) follows from the Neumann series expansion of the resolvent of $L_g^*(t,\theta)$. $\Box$

\bigskip

\noindent {\it Proof of Theorem 4.3.}
\noindent $(i)$ The resolvent formula
\begin{equation}
\label{resolvent} (L^*_g (t,\theta)-z)^{-1} = (\L_0 (\theta
)-z)^{-1}(1+ g\tilde{V}^{tot}(t,\theta )(\L_0 (\theta )-z)^{-1})^{-1} \; ,
\end{equation}
holds for small $g$ and $z$ in the half-plane $\{ z\in
{\mathbf C}: 0< c< \Im z \}$. Note that
\begin{align*}
\| g\tilde{V}^{tot} (t, \theta) (\L_0 (\theta)-z)^{-1} \| &\le |g| C \| (\L_0
(\theta)-z)^{-1} \|\\
&\le |g|C\frac{1}{dist(z,\eta(\L_0(\theta)))},\label{ResEst}
\end{align*}
where $C$ is given by \fer{C} and $\eta(\L_0(\theta))$ is the closure of the numerical range of $\L_0.$ Fix $g_1$ such that it satisfies \fer{g1}, and choose $\epsilon$ such that $\epsilon> \frac{k-|\nu|}{2} >0$. Let
$$G(\nu, \epsilon):= \{ z\in {\mathbf C}: \Im z > \nu
; dist(z,\eta (\L_0 (\theta))>\epsilon\}.$$
Then
\begin{equation*}
\sup_{z\in G(\nu,\epsilon)}\| g \tilde{V}^{tot}(t, \theta)(\L_0
(\theta)-z)^{-1}\| \le \frac{|g|}{g_1},
\end{equation*}
uniformly in $t.$
If $|g| < g_1$, the resolvent formula (\ref{resolvent})
holds on $G(\nu, \epsilon)$, and, for $m\ge 1$,
\begin{equation}
\sup_{z \in G( \nu , \epsilon )} \| (z-L^*_g (t,\theta))^{-1}-\sum_{j=0}^{m-1}(z-\L_0 (t,\theta))^{-1}(g\tilde{V}^{tot}(t ,
\theta)(z-\L_0(\theta))^{-1})^j \|
\le \frac{(\frac{|g|}{g_1})^m}{1-\frac{|g|}{g_1}},
\label{extendedresolvent}
\end{equation}
uniformly in $t.$
It follows that
\begin{equation}
\bigcup_{\epsilon>\frac{k-|\nu|}{2}} G(\nu, \epsilon)\subset \rho (L^*_g(t,\theta)) \; ,
\end{equation}
where $\rho (L^*_g(t,\theta))$ is the resolvent set of $L^*_g (t, \theta)$.
Moreover, setting $m=1$ in \fer{extendedresolvent}, it follows that, for $|g|<g_1/2,$
$$\| P_{g}(t,\theta)-P_{0}\| < 1,$$
and hence $P_{g}(t,\theta)$ is analytic in $g.$ 

Fix $(g_0, \theta_0)\in {\mathbf C}\times I^- (\delta)$
such that $|g_0| < g_1$. Since
$L^*_{g_0}(t,\theta_0)$ and $L^*_{g_0}(t,\theta)$ are unitarily
equivalent if $(\theta - \theta_0)\in {\mathbf R}$ and the discrete eigenvalues of $L_{g_0}^*(t,\theta)$ are analytic functions with at most algebraic singularities in the neighbourhood of $\theta_0$, it follows
that the pure point spectrum of $L^*_{g_0}(t,\theta)$ is independent
of $\theta$.

\noindent$(ii)$ Analyticity of $T_{g}(t)$ in $g$ follows directly from (i) and the definition of $T_{g}(t).$
Since $\| T_{g}(t) - 1 \| < 1$ for $|g|< g_1 /2$,
$T_{g}^{-1/2}(t)$ is also analytic in $g$. Inserting the Neumann series
for the resolvent of $L^*_g(t,\theta)$, gives
\begin{equation}
T_{g}(t) = 1 + \sum_{j=1}^{\infty} g^j T^{(j)}(t) \; ,
\end{equation}
with
\begin{equation}
T^{(j)}(t) =\oint_{\gamma} \frac{dz}{2\pi i} (z-\L^\S)^{-1} P_{0}
\tilde{V}^{tot}(t, \theta) ((z-\L_0(\theta))^{-1}\tilde{V}^{tot}(t,
\theta))^{j-1}P_{0} (z-\L^\S)^{-1}\; .
\end{equation}
Similarly,
\begin{equation}
M_{g}(t)=\L^\S+\sum_{j=1}^{\infty}g^j M^{(j)}(t)\; ,
\end{equation}
with
\begin{equation}
M^{(j)}(t)=\oint_\gamma \frac{dz}{2\pi i}z (z-\L^\S)^{-1} P_{0}
\tilde{V}^{tot}(t, \theta) ((z-\L_0(\theta))^{-1}\tilde{V}^{tot}(t,
\theta))^{j-1}P_{0} (z-\L^\S)^{-1}\; .
\end{equation}
The odd terms in the above two expansions are zero due to the fact
that $P_{0}$ projects onto the $N=0$ sector. The first non-trivial
coefficient in the Taylor series of $\S_{g}(t)$ is
\begin{align}
\S^{(2)}(t) &= M^{(2)}(t)-\frac{1}{2} (T^{(2)}(t) \L^\S + \L^\S T^{(2)}(t)) \\
&= \frac{1}{2} \oint_{\gamma} \frac{dz}{2\pi
i}(\xi(z,t)(z-\L^\S)^{-1}+(z-\L^\S)^{-1}\xi(z,t))\; , 
\label{2orderqL}
\end{align}
with
$$\xi(z,t)=P_{0}\tilde{V}_g^{tot}(t,
\theta)(z-\L_0(\theta))^{-1}\tilde{V}_g^{tot}(t, \theta)P_{0}.$$ $\Box$

\bigskip


\noindent{\it Details of the calculation of the discrete spectrum of $L_g^*(t,\theta)$}
\bigskip

Denote by $P_k, k=0,\cdots ,3,$ the spectral projection onto the eigenstates of $\L^\S,$ and let 
\begin{equation*}
\label{gamma2}
\Gamma^{(2)}_k (t):= P_k \S^{(2)}(t)P_k \; , k=0,\cdots,3.
\end{equation*}
Consider first the nondegenerate eigenvalues ($E_k=\mp 2\omega_0,k=2,3$).
Using the fact that 
\begin{align*}
\lim_{\epsilon\searrow 0}\Re \frac{1}{x-i\epsilon} &= \PV \frac{1}{x} ; \\ 
\lim_{\epsilon\searrow 0}\Im \frac{1}{x-i\epsilon}&= i\pi\delta (x) ,
\end{align*}
and applying the Cauchy integration formula gives
\begin{eqnarray*}
\Re  \Gamma^{(2)}_3 &=& \frac{1}{2}\sum_i \PV \int_{{\mathbf R}} du \frac{\| \tilde{f}_i(u,t)\|^2_\B}{u-2\omega_0}
\; , \\
\Im \Gamma^{(2)}_3 &=& -\frac{\pi}{2} \sum_i \| \tilde{f}_i(2\omega_0,t)\|^2_\B \; ,
\end{eqnarray*}
and
\begin{eqnarray*}
\Re \Gamma^{(2)}_2 &=& -\frac{1}{2}\sum_i \PV \int_{{\mathbf R}} du \frac{\| \tilde{f}_i(u,t)\|^2_\B}{u-2\omega_0} \; , \\
\Im \Gamma^{(2)}_2 &=& -\frac{\pi}{2} \sum_i  \| \tilde{f}_i(2\omega_0,t)\|^2_\B \; .
\end{eqnarray*}

Now apply degenerate perturbation theory for the zero eigenvalue. Using the definition of $f_{i,\beta_i}$ and $f^\#_{i,\beta_i}$ given in section 3, 
\begin{eqnarray*}
\Re \Gamma^{(2)}_{0,1} &=& 0 \; , \\
\Im \Gamma^{(2)}_{0,1} &=& -\pi \sum_i 
\frac{\|\tilde{f}_i(2\omega_0,t)\|^2_\B}{cosh(\beta_i\omega_0)}
\left(
\begin{matrix}
e^{\beta_i \omega_0} & -e^{\beta_i\omega_0} \\
-e^{-\beta_i\omega_0} & e^{-\beta_i\omega_0} 
\end{matrix}
\right) \; .
\end{eqnarray*}
Therefore, to second order in the coupling $g$,
\begin{align*}
E_{2,3}(g,t)=&\mp (2\omega_0 -\frac{1}{2} g^2 \PV\int_{\mathbf R} du \frac{1}{2\omega_0-u}
\sum_i \| \tilde{f}_i(u,t)\|^2_{\B})\\& -i\frac{\pi}{2} g^2
\sum_i \|\tilde{f}_i(2\omega_0,t)\|^2_\B + O(g^4) \; ,
\end{align*}
while
\begin{equation*}
E_{0,1}(g,t)= g^2 a_{0,1}(t) + O(g^4) \; ,
\end{equation*}
where $a_{0,1}(t)$ are the eigenvalues of the matrix
\begin{equation*}
-i\pi \sum_i 
\frac{\|\tilde{f}_i(2\omega_0,t)\|^2_\B}{2\cosh(\beta_i\omega_0)}
\left(
\begin{matrix}
e^{\beta_i \omega_0} & -e^{\beta_i\omega_0} \\
-e^{-\beta_i\omega_0} & e^{-\beta_i \omega_0} 
\end{matrix}
\right) \; .
\end{equation*}

Since $\Omega$ is an eigenvector corresponding to the isolated
zero eigenvalue of $L_g(t,\theta)$ (by construction,
$L_g(t, \theta)\Omega=0$), then zero is also an
isolated eigenvalue of $L_g^*(t, \theta)$. (One way of seeing this
is to take the adjoint of the spectral projection of $L_g(t, \theta)$
corresponding to $\Omega$, which is defined using contour integration.) Note that $\psi=\left(\begin{matrix} 1
\\ 1 
\end{matrix}\right)$ is the eigenvector corresponding to the
zero eigenvalue of $\S^{(2)}(t)$. Hence,
\begin{align*}
E_0(g,t)&=0 \; ,\\
E_1(g,t)&=-i\pi g^2 \sum_i \| \tilde{f}_i (2\omega_0,t) \|^2_\B
+O(g^4)\; .
\end{align*}

\bigskip

\noindent {\it Proof of Corollary 4.4 (NESS)}

{\it Proof.} Define $k:=\min(\frac{\pi}{\beta_1},
\cdots,\frac{\pi}{\beta_n},\delta)$, where $\delta$ appears in assumption
(B2), and let $\theta\in I^-(k)$. We already know the
spectrum of $L_g^*(t,\theta)$ from Theorem 4.3.  For $a\in
\O^{test}$, 
\begin{align*}
\lim_{t\rightarrow\infty} \langle \Omega , \a^t_g (a) \Omega
\rangle &= \lim_{t\rightarrow\infty}\langle \Omega, e^{itL_g}a
e^{-itL_g}\Omega \rangle \\
&= \lim_{t\rightarrow\infty} \langle e^{-itL_g^*} \Omega , a
\Omega
\rangle \\
&= \lim_{t\rightarrow\infty} \langle e^{-itL_g^*(\theta)} \Omega ,
a(\overline{\theta}) \Omega
\rangle \\
&= \lim_{t\rightarrow\infty} \frac{1}{2\pi i} \langle
\int_{-\infty}^\infty du (u+i\eta -
L_g^*(\theta))^{-1}e^{-i(u+i\eta)t} \Omega ,a (\overline{\theta})
\Omega\rangle \; ,
\end{align*}
for $\eta>0$. One may decompose the last integral into two parts (see for example [JP1]). The first part is
\begin{equation*}
\lim_{t\rightarrow\infty} \frac{1}{2\pi i} \langle
\oint_{\gamma}dz (z -L_g^*(\theta ) )^{-1} e^{-izt}\Omega,
a(\overline{\theta}) \Omega \rangle = \langle \Omega_g , D^{-1} a
\Omega \rangle \; ,
\end{equation*}
where the zero-energy resonance is
\begin{equation*}
\Omega_g:=DU(-\theta)P^0_g(\theta)U(\theta)D\Omega=
DU(-\theta)P^0_g(\theta)\Omega \; .
\end{equation*}
The second term converges to zero exponentially fast as
$t\rightarrow\infty$, since
\begin{equation*}
\frac{1}{2\pi i} \langle \int_{-\infty}^\infty
(u-i(\mu-\epsilon)-L_g^*(\theta))^{-1}
e^{-i(u-i(\mu-\epsilon))t}\Omega, a(\overline{\theta}) \Omega
\rangle = O(e^{-(\mu-\epsilon')t}) \; ,
\end{equation*}
where $0<\epsilon'<\epsilon<|\Im\theta|=:\mu$; (see also Theorem 19.2 in [Rud]).$\Box$

\bigskip


\end{document}